# COHERENT CURRENT STATES IN A TWO-BAND SUPERCONDUCTOR.

**Y.S.Yerin, A.N.Omelyanchouk**

B.Verkin Istitute for Low Temperature Physics and Engineering of the National Academy of Science of Ukraine, 47 Lenin Ave., Kharkov, 61103, Ukraine

E-mail: omelyanchouk@ilt.kharkov.ua

**ABSTRACT**

Homogeneous current states in thin films and Josephson current in superconducting microbridges are studied within the frame of a two-band Ginzburg-Landau theory. By solving the coupled system of equations for two order parameters the depairing current curves and Josephson current-phase relation are calculated for different values of phenomenological parameters $\gamma$ and $\eta$. Coefficients $\gamma$ and $\eta$ describe the coupling of order parameters (proximity effect) and their gradients (drag effect) respectively. For definite values of parameters the dependence of current $j$ on superfluid momentum $q$ contains local minimum and corresponding bi-stable states. It is shown that the Josephson microbridge from two-band superconductor can demonstrate $\pi$-junction behaviour.

Key words: Current states, two-band superconductivity, proximity effect, drag effect, Josephson microbridge, $\pi$-junction.

# 1. INTRODUCTION

To present day overwhelming majority works on theory of superconductivity were devoted to single gap superconductors. More than 40 years ago the possibility of superconductors with two superconducting order parameters were considered by V. Moskalenko [1] and H. Suhl, B.Matthias and L.Walker [2]. In the model of superconductor with the overlapping energy bands on Fermi surface V.Moskalenko has theoretically investigated the thermodynamic and electromagnetic properties of two-band superconductors. The real boom in investigation of multi-gap superconductivity started after the discovery of two gaps in $MgB_2$ [3] by the scanning tunneling [4, 5] and point contact spectroscopy [6, 7, 8]. The compound $MgB_2$ has the highest critical temperature $T_c = 39$ K among superconductors with phonon mechanism of the pairing and two energy gaps $\Delta_1 \approx 7 meV$ and $\Delta_2 \approx 2,5 meV$ at $T = 0$. At this time two-band superconductivity is studied also in another systems, *e.g.* in heavy fermion compounds [9,10], borocarbides [11] and liquid metallic hydrogen [12-14]. Various thermodynamic and transport properties of $MgB_2$ were studied in the framework of two-band BCS model [15-22]. Ginzburg-Landau functional for two-gap superconductors was derived within the weak-coupling BCS theory in dirty [23] and clean [24] superconductors. Within the Ginzburg-Landau scheme the magnetic properties [25, 26, 27] and peculiar vortices [28, 29, 30] were studied.

The aim of this article is to present Ginzburg-Landau theory of the current carrying states in superconductors with two order parameters. In the case of several order parameters the qualitatively new features in superconducting current state are related to mutual influence of the modules of complex order parameters as well of the gradients of their phases. We study the manifestations of these effects in the current–momentum dependence and in the Josephson current-phase relation. In Section 2 the general phenomenological description of two-band superconductors within Ginzburg-Landau theory without external magnetic field is given. The Ginzburg-Landau equations for two coupled superconducting order parameters include the proximity and drag effects. In Section 3 the peculiarities of homogeneous current states in multi-gap superconductors are studied. The dependence of current on superfluid momentum for different values of parameters is calculated. We demonstrate that for definite values of parameters it contains local minima and corresponding bi-stable states in GL free energy. In Section 4 the Josephson effect in simple model of weak superconducting link (generalization of Aslamazov-Larkin theory [31] on two-band superconductor) is considered and possibility of $\pi$-junction behaviour is demonstrated.

## 2. GINZBURG-LANDAU EQUATIONS FOR TWO-BAND SUPERCONDUCTIVITY

The phenomenological Ginzburg-Landau (GL) free energy density functional for two coupled superconducting order parameters $\psi_1$ and $\psi_2$ can be written as

$$F_{GL} = F_1 + F_2 + F_{12} + \frac{(\text{rot}\vec{A})^2}{8\pi},$$

where

$$F_1 = \alpha_1 |\psi_1|^2 + \frac{1}{2}\beta_1 |\psi_1|^4 + \frac{1}{2m_1}\left|\left(-i\hbar\nabla - \frac{2e}{c}\vec{A}\right)\psi_1\right|^2 \quad (1)$$

$$F_2 = \alpha_2 |\psi_2|^2 + \frac{1}{2}\beta_2 |\psi_2|^4 + \frac{1}{2m_2}\left|\left(-i\hbar\nabla - \frac{2e}{c}\vec{A}\right)\psi_2\right|^2 \quad (2)$$

and

$$F_{12} = -\gamma\left(\psi_1^*\psi_2 + \psi_1\psi_2^*\right) + \eta\left(\left(-i\hbar\nabla - \frac{2e}{c}\vec{A}\right)\psi_1\left(i\hbar\nabla - \frac{2e}{c}\vec{A}\right)\psi_2^* + \left(i\hbar\nabla - \frac{2e}{c}\vec{A}\right)\psi_1^*\left(-i\hbar\nabla - \frac{2e}{c}\vec{A}\right)\psi_2\right) \quad (3)$$

The terms $F_1$ and $F_2$ are conventional contributions from $\psi_1$ and $\psi_2$, term $F_{12}$ describes without the loss of generality the interband coupling of order parameters. The coefficients $\gamma$ and $\eta$ describe the coupling of two order parameters (proximity effect) and their gradients (drag effect) [25-27], respectively.

By minimization of the free energy $F=\int (F_1+F_2+F_{12}+\frac{H^2}{8\pi})d^3r$ with respect to $\psi_1$, $\psi_2$ and $\vec{A}$ we obtain the differential GL equations for two-band superconductor

$$\begin{cases} \frac{1}{2m_1}\left(-i\hbar\nabla-\frac{2e}{c}\vec{A}\right)^2\psi_1+\alpha_1\psi_1+\beta_1|\psi_1|^2\psi_1-\gamma\psi_2+\eta\left(-i\hbar\nabla-\frac{2e}{c}\vec{A}\right)^2\psi_2=0 \\ \frac{1}{2m_2}\left(-i\hbar\nabla-\frac{2e}{c}\vec{A}\right)^2\psi_2+\alpha_2\psi_2+\beta_2|\psi_2|^2\psi_2-\gamma\psi_1+\eta\left(-i\hbar\nabla-\frac{2e}{c}\vec{A}\right)^2\psi_1=0 \end{cases} \quad (4)$$

and expression for the supercurrent

$$\begin{aligned} \vec{j}=&-\frac{ie\hbar}{m_1}\left(\psi_1^*\nabla\psi_1-\psi_1\nabla\psi_1^*\right)-\frac{ie\hbar}{m_2}\left(\psi_2^*\nabla\psi_2-\psi_2\nabla\psi_2^*\right)-\\ &-2ie\hbar\eta\left(\psi_1^*\nabla\psi_2-\psi_2\nabla\psi_1^*-\psi_1\nabla\psi_2^*+\psi_2^*\nabla\psi_1\right)-\\ &-\left(\frac{4e^2}{m_1c}|\psi_1|^2+\frac{4e^2}{m_2c}|\psi_2|^2+\frac{8\eta e^2}{c}\left(\psi_1^*\psi_2+\psi_2^*\psi_1\right)\right)\vec{A} \end{aligned} \quad (5)$$

In the absence of currents and gradients of order parameters modules the equilibrium values of order parameters $\psi_{1,2}=\psi_{1,2}^{(0)}e^{i\chi_{1,2}}$ are determined by the set of coupled equations

$$\begin{aligned} \alpha_1\psi_1^{(0)}+\beta_1\psi_1^{(0)3}-\gamma\psi_2^{(0)}e^{i(\chi_2-\chi_1)}&=0,\\ \alpha_2\psi_2^{(0)}+\beta_2\psi_2^{(0)3}-\gamma\psi_1^{(0)}e^{i(\chi_1-\chi_2)}&=0. \end{aligned} \quad (6)$$

For the case of two order parameters the question arises about the phase difference $\phi = \chi_1 - \chi_2$ between $\psi_1$ and $\psi_2$. In homogeneous no-current state, by analyzing the free energy term $F_{12}$ (3), one can obtain that for $\gamma > 0$ phase shift $\phi = 0$ and for $\gamma < 0$ $\phi = \pi$. The statement, that $\phi$ can have only values 0 or $\pi$ takes place also in a current carrying state, but for coefficient $\eta \neq 0$ the criterion for $\phi$ equals 0 or $\pi$ depends now on the value of the current (see below).

If the interband interaction is ignored, the equations (4) are decoupled into two ordinary G-L equations with two different critical temperatures $T_{c_1}$ and $T_{c_2}$. In general, independently of the sign of $\gamma$, the superconducting phase transition results at a well-defined temperature exceeding both $T_{c_1}$ and $T_{c_2}$, which is determined from the equation:

$$\alpha_1(T_c)\alpha_2(T_c) = \gamma^2 \tag{7}$$

Let the first order parameter is stronger then second one, i.e. $T_{c_1} > T_{c_2}$. Following [24] we represent temperature dependent coefficients as

$$\begin{aligned}\alpha_1(T) &= -a_1(1 - T/T_{c1}), \\ \alpha_2(T) &= a_{20} - a_2(1 - T/T_{c1}).\end{aligned} \tag{8}$$

Phenomenological constants $a_{1,2}, a_{20}$ and $\beta_{1,2}, \gamma$ can be related to microscopic parameters in two-band BCS model. From (7) and (8) we obtain for critical temperature $T_c$:

$$T_c = T_{c1}\left(1 + \sqrt{\left(\frac{a_{20}}{2a_2}\right)^2 + \frac{\gamma^2}{a_1 a_2}} - \frac{a_{20}}{2a_2}\right). \tag{9}$$

For arbitrary value of the interband coupling $\gamma$ Eq.(6) can be solved numerically. For $\gamma = 0$, $T_c = T_{c1}$ and for temperature close to $T_c$ (hence for $T_{c2} < T \leq T_c$) equilibrium values of the order parameters are $\psi_2^{(0)}(T) = 0$, $\psi_1^{(0)}(T) = \sqrt{a_1(1 - T/T_c)/\beta_1}$. Considering in the following weak interband coupling, we have from Eqs. (6-9) corrections $\sim \gamma^2$ to these values:

$$\psi_1^{(0)}(T)^2 = \frac{a_1}{\beta_1}(1 - \frac{T}{T_c}) + \frac{\gamma^2}{\beta_1}\left(\frac{1}{a_{20} - a_2(1 - \frac{T}{T_c})} - \frac{T}{T_c}\frac{1}{a_{20}}\right),$$

$$\psi_2^{(0)}(T)^2 = \frac{a_1}{\beta_1}(1 - \frac{T}{T_c})\frac{\gamma^2}{(a_{20} - a_2(1 - \frac{T}{T_c}))^2}.$$

(10)

Expanding expressions (9) over $(1 - \frac{T}{T_c}) \ll 1$ we have conventional temperature dependence of equilibrium order parameters in weak interband coupling limit

$$\psi_1^{(0)}(T) \approx \sqrt{\frac{a_1}{\beta_1}}\left(1 + \frac{1}{2}\frac{a_{20} + a_2}{a_{20}^2 a_1}\gamma^2\right)\sqrt{1 - \frac{T}{T_c}},$$

$$\psi_2^{(0)}(T) \approx \sqrt{\frac{a_1}{\beta_1}}\frac{\gamma}{a_{20}}\sqrt{1 - \frac{T}{T_c}}.$$

(11)

Considered above case (expressions (9)-(11)) corresponds to different critical temperatures $T_{c_1} > T_{c_2}$ in the absence of interband coupling $\gamma$. Order parameter in the second band $\psi_2^{(0)}$ arises from the "proximity effect" of stronger $\psi_1^{(0)}$ and is proportional to value of $\gamma$ (11). Consider now another situation. Suppose for simplicity that two condensates in current zero state are identical. In this case for arbitrary value of $\gamma$ we have

$$\alpha_1(T) = \alpha_2(T) \equiv \alpha(T) = -a\left(1 - \frac{T}{T_c}\right), \beta_1 = \beta_2 \equiv \beta. \tag{12}$$

$$\psi_1^{(0)} = \psi_2^{(0)} = \sqrt{\frac{|\gamma| - \alpha}{\beta}}. \tag{13}$$

## 3. HOMOGENEOUS CURRENT STATES AND GL DEPAIRING CURRENT

In this section we will consider the homogeneous current states in thin wire or film with transverse dimension $d \ll \xi_{1,2}(T), \lambda_{1,2}(T)$ (see Fig.1), where $\xi_{1,2}(T)$ and $\lambda_{1,2}(T)$ are coherence lengths and London penetration depths for each order parameter correspondingly without interband interaction. This condition leads to one-dimensional problem and permits us to neglect self-magnetic field of the system.

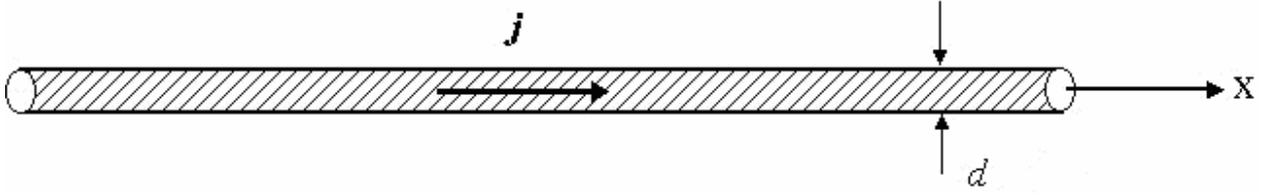

Fig.1. Geometry of the system.

The current density $j$ and modules of the order parameters do not depend on the longitudinal direction $x$. Writing $\psi_{1,2}(x)$ as $\psi_{1,2} = |\psi_{1,2}| \exp(i\chi_{1,2}(x))$ and introducing the difference and weighted sum phases:

$$\begin{cases} \phi = \chi_1 - \chi_2, \\ \theta = c_1\chi_1 + c_2\chi_2, \end{cases}$$

for the free energy density (1)-(3) obtain

$$F = \alpha_1|\psi_1|^2 + \alpha_2|\psi_2|^2 + \frac{1}{2}\beta_1|\psi_1|^4 + \frac{1}{2}\beta_2|\psi_2|^4 + \hbar^2\left(\frac{|\psi_1|^2}{2m_1} + \frac{|\psi_2|^2}{2m_2} + 2\eta|\psi_1||\psi_2|\cos\phi\right)\left(\frac{d\theta}{dx}\right)^2 +$$
$$+\hbar^2\left(c_2^2\frac{|\psi_1|^2}{2m_1} + c_1^2\frac{|\psi_2|^2}{2m_2} - 2\eta c_1 c_2|\psi_1||\psi_2|\cos\phi\right)\left(\frac{d\phi}{dx}\right)^2 - 2\gamma|\psi_1||\psi_2|\cos\phi \quad (14)$$

Where

$$c_1 = \frac{\dfrac{|\psi_1|^2}{m_1} + 2\eta|\psi_1||\psi_2|\cos\phi}{\dfrac{|\psi_1|^2}{m_1} + \dfrac{|\psi_2|^2}{m_2} + 4\eta|\psi_1||\psi_2|\cos\phi}, \quad c_2 = \frac{\dfrac{|\psi_2|^2}{m_2} + 2\eta|\psi_1||\psi_2|\cos\phi}{\dfrac{|\psi_1|^2}{m_1} + \dfrac{|\psi_2|^2}{m_2} + 4\eta|\psi_1||\psi_2|\cos\phi}.$$ (15)

The current density $j$ in terms of phases $\theta$ and $\phi$ has the following form

$$j = 2e\hbar\left(\frac{|\psi_1|^2}{m_1} + \frac{|\psi_2|^2}{m_2} + 4\eta|\psi_1||\psi_2|\cos\phi\right)\frac{d\theta}{dx}$$ (16)

and includes the partial inputs $j_{1,2}$ and proportional to $\eta$ the drag current $j_{12}$.

In contrast to the case of single order parameter [32], the condition $\mathrm{div}\,\mathbf{j} = 0$ does not fix the constancy of superfluid velocity. In appendix we present the Euler – Lagrange equations for $\theta(x)$ and $\phi(x)$. They are complicated coupled nonlinear equations, which generally permit the soliton like solutions (in the case $\eta = 0$ they were considered in [33]). The possibility of states with inhomogeneous phase $\phi(x)$ is needed in separate investigation. Here, we restrict our consideration by the homogeneous phase difference between order parameters $\phi = const$. For $\phi = const$ from equations (A8) (see Appendix) follows that $\theta(x) = qx$ (q is total superfluid momentum) and $\sin\phi = 0$, i.e. $\phi$ equals 0 or $\pi$. Minimization of free energy for $\phi$ gives

$$\cos\phi = sign(\gamma - \eta\hbar^2 q^2)$$ (17)

Note, that now the value of $\phi$, in principle, depends on q, thus, on current density $j$.

Finally, the expressions (14), (16) take the form:

$$F = \alpha_1|\psi_1|^2 + \frac{1}{2}\beta_1|\psi_1|^4 + \frac{\hbar^2}{2m_1}|\psi_1|^2 q^2 + \alpha_2|\psi_2|^2 + \frac{1}{2}\beta_2|\psi_2|^4 + \frac{\hbar^2}{2m_2}|\psi_2|^2 q^2 - 2(\gamma - \eta\hbar^2 q^2)|\psi_1||\psi_2|sign(\gamma - \eta\hbar^2 q^2),$$ (18)

$$j = 2e\hbar \left( \frac{|\psi_1|^2}{m_1} + \frac{|\psi_2|^2}{m_2} + 4\eta |\psi_1||\psi_2| \, sign(\gamma - \eta \hbar^2 q^2) \right) q \tag{19}$$

We will parameterize the current states by the value of superfluid momentum $q$, which for given value of $j$ is determined by equation (19). The dependence of the order parameter modules on $q$ determines by GL equations:

$$\alpha_1 |\psi_1| + \beta_1 |\psi_1|^3 + \frac{\hbar^2}{2m_1} |\psi_1| q^2 - |\psi_2|(\gamma - \eta \hbar^2 q^2) \, sign(\gamma - \eta \hbar^2 q^2) = 0, \tag{20}$$

$$\alpha_2 |\psi_2| + \beta_2 |\psi_2|^3 + \frac{\hbar^2}{2m_2} |\psi_2| q^2 - |\psi_1|(\gamma - \eta \hbar^2 q^2) \, sign(\gamma - \eta \hbar^2 q^2) = 0 \tag{21}$$

At the beginning we consider the case of small values of interband coupling $\gamma$ and dragging coefficient $\eta$. In the same manner as for $q = 0$ (section 2) instead expression (11), for $|\psi_1|(q)$ and $|\psi_2|(q)$ we obtain:

$$\psi_1^2(q) = \frac{a_1 \left(1 - \frac{T}{T_c}\right)}{\beta_1} - \frac{\hbar^2}{2m_1 \beta_1} q^2 - \frac{\gamma^2}{a_{20} \beta_1} \frac{T}{T_c} + \frac{(\gamma - \eta \hbar^2 q^2)^2}{\beta_1 \left(a_{20} - a_2 \left(1 - \frac{T}{T_c}\right) + \frac{\hbar^2}{2m_2} q^2\right)}, \tag{22}$$

$$\psi_1^2(q) = \left( \frac{a_1 \left(1 - \frac{T}{T_c}\right)}{\beta_1} - \frac{\hbar^2}{2m_1 \beta_1} q^2 \right) \frac{(\gamma - \eta \hbar^2 q^2)^2}{\beta_1 \left(a_{20} - a_2 \left(1 - \frac{T}{T_c}\right) + \frac{\hbar^2}{2m_2} q^2\right)}. \tag{23}$$

The system of equations (19), (22), (23) describes the depairing curve $j(q, T)$ and the dependences $|\psi_1|$ and $|\psi_2|$ on the current $j$ and the temperature $T$. It can be solved numerically for given superconductor with concrete values of phenomenological parameters.

In order to study the specific effects produced by interband coupling and dragging consider now the model case when order parameters coincide at $j = 0$ (eqs. (12), (13)) but gradient terms in equations (4) are different. Eqs. (19)-(21) in this case take the form

$$f_1 \left(1 - (1 + \tilde{\gamma}) f_1^2\right) - f_1 q^2 + f_2 (\tilde{\gamma} - \tilde{\eta} q^2) \, sign(\tilde{\gamma} - \tilde{\eta} q^2) = 0 \tag{24}$$

$$f_2\left(1-(1+\tilde{\gamma})f_2^2\right)-kf_2q^2+f_1\left(\tilde{\gamma}-\tilde{\eta}q^2\right)sign\left(\tilde{\gamma}-\tilde{\eta}q^2\right)=0 \qquad (25)$$

$$j=f_1^2q+kf_2^2q+2\tilde{\eta}f_1f_2q\,sign\left(\tilde{\gamma}-\tilde{\eta}q^2\right) \qquad (26)$$

Here we normalize $\psi_{1,2}$ on the value of the order parameters at $j=0$ (13), $j$ is measured in units of $2\sqrt{2}e\dfrac{|\gamma|+|\alpha|}{\beta}\sqrt{\dfrac{|\alpha|}{m_1}}$, $q$ is measured in units of $\sqrt{\dfrac{\hbar^2}{2m_1|\alpha|}}$, $\tilde{\gamma}=\dfrac{\gamma}{|\alpha|}$, $\tilde{\eta}=2\eta m_1$, $k=\dfrac{m_1}{m_2}$. If $k=1$ order parameters coincides also in current-carrying state $f_1=f_2=f$ and from eqs. (24)-(26) we have the expressions

$$f^2(q)=\dfrac{1-q^2-|\tilde{\gamma}-\tilde{\eta}q^2|}{1+|\tilde{\gamma}|}, \qquad (27)$$

$$j(q)=2f^2\left(1-\tilde{\eta}sign\left(\tilde{\gamma}-\tilde{\eta}q^2\right)\right)q, \qquad (28)$$

which for $\tilde{\gamma}=\tilde{\eta}=0$ are conventional dependences for one-band superconductor [32] (see Fig. 2 a,b).

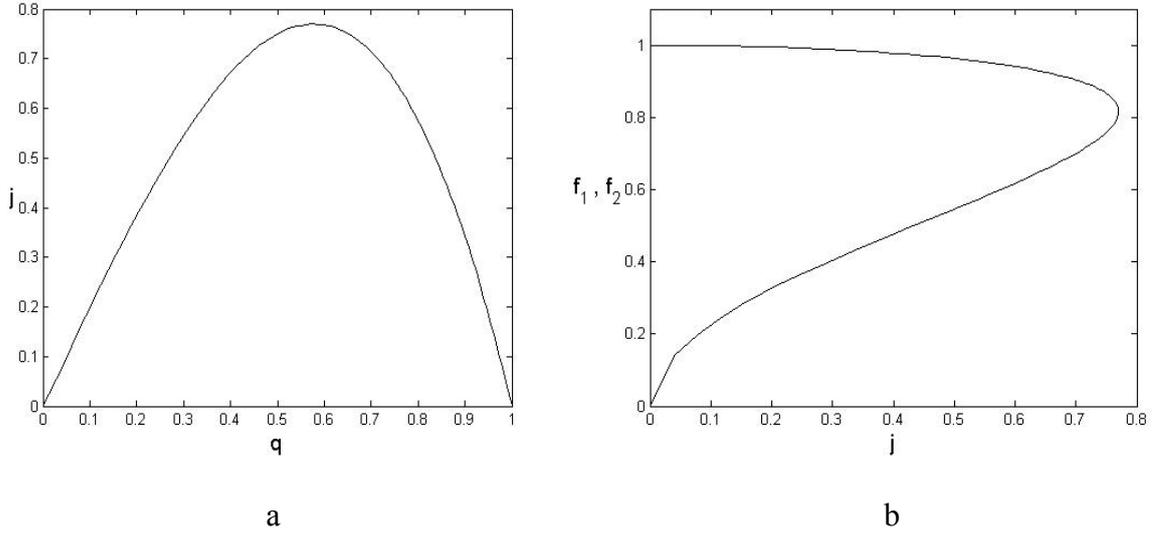

a  b

Fig.2 a,b. Depairing current curve (a) and dependence of the order parameter modules vs. current (b) for coincident order parameters.

For $k \neq 1$ depairing curve $j(q)$ can contain two increasing with $q$ stable branches, which corresponds to possibility of bistable state. In Fig. 3 the numerically calculated from equations (24-26) the curve $j(q)$ and dependences $f_1(j), f_2(j)$ are shown for $k=5$ and $\tilde{\gamma} = \tilde{\eta} = 0$.

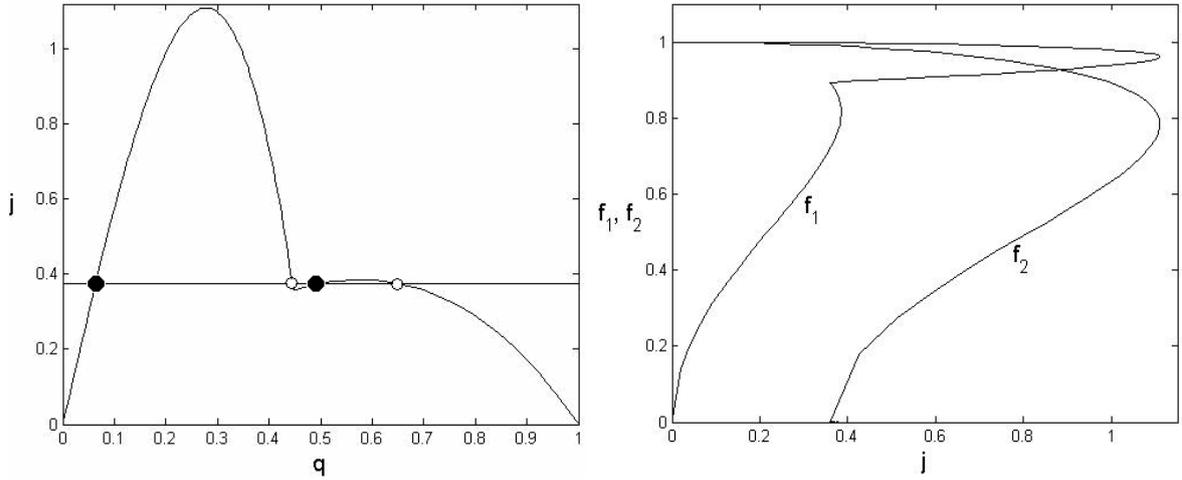

Fig.3. (a) Dependence of the current $j$ on superfluid momentum $q$. For value of the current $j=j_0$ stable states (●) and unstable states (○) are shown.( b) Dependences of the order parameters on current $j$, $k=5$ and $\tilde{\gamma} = \tilde{\eta} = 0$.

The interband scattering ($\tilde{\gamma} = 0$) smears the second peak in $j(q)$, see Fig.4.

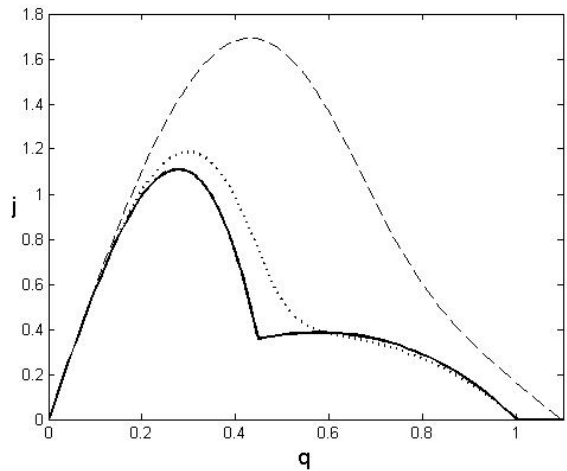

Fig.4. Depairing current curves for different values of the interband interaction: $\tilde{\gamma} = 0$ (solid line), $\tilde{\gamma} = 0.1$ (dotted line) and $\tilde{\gamma} = 1$ (dashed line). Ratio of the effective masses equals $k = 5$, $\tilde{\eta} = 0$.

If dragging effect ($\tilde{\eta} = 0$) is taking into account the depairing curve $j(q)$ can contain the jump at definite value of $q$ (for $k = 1$ see eq. 28), see Fig.5. This jump corresponds to the switching of relative phase difference from 0 to π.

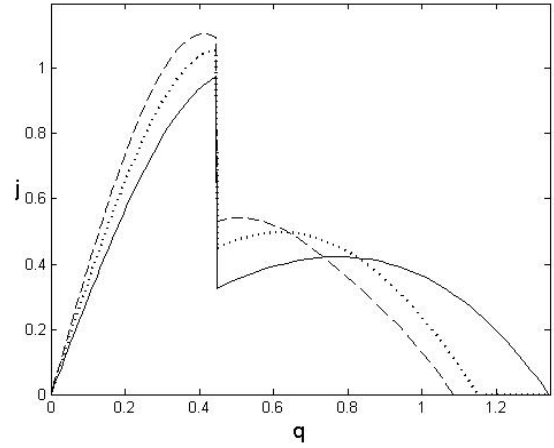

Fig.5. Depairing current curves for different values of the effective masses ratio $k = 1$ (solid line), $k = 1.5$ (dotted line) and $k = 2$ (dashed line). Interband interaction coefficient $\tilde{\gamma} = 0.1$ and drag effect coefficient $\tilde{\eta} = 0.5$.

# 4. JOSEPHSON EFFECT IN TWO-BAND SUPERCONDUCTING MICROCONSTRICTION

In the previous section GL-theory of two-band superconductors was applied for filament's length $L \to \infty$. Opposite case of the strongly inhomogeneous current state is the Josephson microbridge geometry, which we model as narrow channel connecting two massive superconductors (banks). The length $L$ and the diameter $d$ of the channel (see Fig. 6) are assumed to be small as compared to the order parameters coherence lengths $\xi_1, \xi_2$.

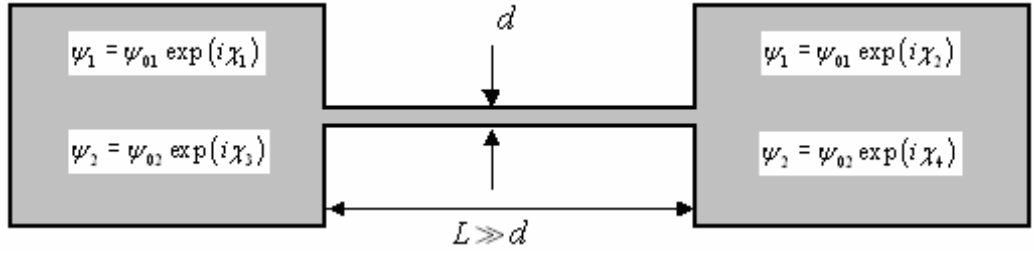

Fig.6. Geometry of S-C-S contact as narrow superconducting channel in contact with bulk two-band superconductors. The values of the order parameters at the banks are indicated.

For $d \ll L$ we can solve one-dimensional GL equations (4) inside the channel with the rigid boundary conditions for order parameters at the ends of the channel [34].

In the case $L \ll \xi_1, \xi_2$ we can neglect in equations (4) all terms except the gradient ones and solve equations:

$$\begin{cases} \dfrac{d^2 \psi_1}{dx^2} = 0, \\ \dfrac{d^2 \psi_2}{dx^2} = 0 \end{cases} \tag{29}$$

with the boundary conditions:

$$\psi_1(0) = \psi_{01} \exp(i\chi_1), \psi_2(0) = \psi_{02} \exp(i\chi_3),$$

$$\psi_1(L) = \psi_{01}\exp(i\chi_2), \psi_2(L) = \psi_{02}\exp(i\chi_4). \tag{30}$$

Calculating the current density $j$ in the channel we obtain:

$$j = j_1 + j_2 + j_{12}, \tag{31}$$

$$j_1 = \frac{2e\hbar}{Lm_1}\psi_{01}^2 \sin(\chi_2 - \chi_1), \tag{32}$$

$$j_2 = \frac{2e\hbar}{Lm_2}\psi_{02}^2 \sin(\chi_4 - \chi_3), \tag{33}$$

$$j_{12} = \frac{4e\hbar}{L}\eta\psi_{01}\psi_{02}\left(\sin(\chi_2 - \chi_3) + \sin(\chi_4 - \chi_1)\right). \tag{34}$$

Let $\chi_2 - \chi_1 = \chi$. The difference between two order parameter phases in the banks equals 0 or $\pi$, depending on the sign of the constant interband interaction $\gamma$. Therefore, if $\gamma > 0$ $\chi_3 = \chi_1$ and $\chi_4 = \chi_2$ and if $\gamma < 0$ then $\chi_3 - \chi_1 = \pi$, $\chi_4 - \chi_2 = \pi$. Thus the current-phase relation $j(\chi)$ in general case of arbitrary values of phenomenological constants $\gamma$ and $\eta$ for two-band superconducting microbridge has the form:

$$j \equiv j_0 \sin\chi = \frac{2e\hbar}{L}\left(\frac{\psi_{01}^2}{m_1} + \frac{\psi_{02}^2}{m_2} + 4\eta\psi_{01}\psi_{02}sign(\gamma)\right)\sin\chi. \tag{35}$$

The value of $j_0$ in (35) can be both positive and negative:

$$j_0 > 0 \text{ if } \eta sign(\gamma) > -\left(\frac{\psi_{01}}{4m_1} + \frac{\psi_{02}}{4m_2}\right), \tag{36}$$

$$j_0 < 0 \text{ if } \eta sign(\gamma) < -\left(\frac{\psi_{01}}{4m_1} + \frac{\psi_{02}}{4m_2}\right). \tag{37}$$

When the condition (37) for set of parameters for two-band superconductor is satisfied the microbridge behaves as the so-called $\pi-$ junction (see *e.g.* review [35]).

## 5. CONCLUSIONS

We have investigated the current carrying states in two-band superconductors within phenomenological Ginzburg-Landau theory. Two limiting situations were considered, homogeneous current state in long film or channel and Josephson effect in short superconducting microconstriction. We used the GL functional for two order parameters which includes the interband coupling (proximity effect) and the effect of dragging in current state of two-band system. For the case of two order parameters the question arises about the phase difference $\phi = \chi_1 - \chi_2$ between $\psi_1 = |\psi_1| e^{i\chi_1}$ and $\psi_2 = |\psi_2| e^{i\chi_2}$. In homogeneous no-current state the value of $\phi$ equals to 0 or π depending on the sign of interband coupling constant $\gamma$ [36]. The statement, that $\phi$ can have only values 0 or $\pi$ takes place also in a current carrying state, but for nonzero drag coefficient $\eta$ the criterion for $\phi$ equals 0 or $\pi$ depends now on the value of the superfluid momentum $q$, namely $\cos\phi = sign(\gamma - \eta\hbar^2 q^2)$. The system of coupled GL equations is analyzed for different values of phenomenological parameters. The depairing current expression contains the term $\cos\phi$ and, in general, depending on parameters γ and η the increasing of momentum $q$ can switch the value of $\phi$ from 0 to $\pi$. In current driven regime it leads to existence of two growing branches of *j(q)*, which both are stable. This bistability is intrinsic property of two-band superconductor. It is interesting to study the effects of relative phase switching in magnetic flux driven regime in multivalued geometry. The Josephson current-phase relation for two band superconducting weak link *j(χ)* also contains the difference of order parameters phases $\phi$ in the banks, $j = j_0(\phi)\sin\chi$. The value of $j_0$ may be as positive as negative. In the last case we have what is called the π-junction, again due to intrinsic properties of two-band superconductivity. In Section 2 we restrict our consideration by the homogeneous phase difference between two order parameters $\phi$. The general equations (A8) permit the possibility of inhomogeneous, soliton-like distributions $\phi(x)$, which will be subject of separate publication.

The authors would like to acknowledge S.V. Kuplevakhsky for useful discussions.

# APPENDIX

## Free energy transformation.

Instead of the phases $\chi_1$ and $\chi_2$ introduce new variables $\phi$ and $\theta$:

$$\begin{cases} \chi_1 - \chi_2 = \phi, \\ c_1\chi_1 + c_2\chi_2 = \theta, \end{cases} \tag{A1}$$

where coefficients $c_1$ and $c_2$ are chosen as

$$c_1 = \frac{\dfrac{|\psi_1|^2}{m_1} + 2\eta|\psi_1||\psi_2|\cos\phi}{\dfrac{|\psi_1|^2}{m_1} + \dfrac{|\psi_2|^2}{m_2} + 4\eta|\psi_1||\psi_2|\cos\phi}, \quad c_2 = \frac{\dfrac{|\psi_2|^2}{m_2} + 2\eta|\psi_1||\psi_2|\cos\phi}{\dfrac{|\psi_1|^2}{m_1} + \dfrac{|\psi_2|^2}{m_2} + 4\eta|\psi_1||\psi_2|\cos\phi} \tag{A2}$$

Expression for the free energy density in new variables takes a quadratic form on derivatives of $\theta$ and $\varphi$:

$$F = A + B\left(\frac{d\theta}{dx}\right)^2 + C\left(\frac{d\phi}{dx}\right)^2 - D\cos\phi \tag{A3}$$

Here $A, B, C, D$ are:

$$A = \alpha_1|\psi_1|^2 + \frac{1}{2}\beta_1|\psi_1|^4 + \alpha_2|\psi_2|^2 + \frac{1}{2}\beta_2|\psi_2|^4 \tag{A4}$$

$$B = \left(\frac{|\psi_1|^2}{2m_1} + \frac{|\psi_2|^2}{2m_2} + 2\eta|\psi_1||\psi_2|\cos\phi\right)\hbar^2 \tag{A5}$$

$$C = \left(c_2^2\frac{|\psi_1|^2}{2m_1} + c_1^2\frac{|\psi_2|^2}{2m_2} - 2c_1c_2\eta|\psi_1||\psi_2|\cos\phi\right)\hbar^2 \tag{A6}$$

$$D = 2\gamma|\psi_1||\psi_2| \tag{A7}$$

Making variation for $\theta$ and $\phi$ we obtain equations for spatial dependence of phases $\phi$ and $\theta$:

$$\begin{cases} \left(\dfrac{|\psi_1|^2}{2m_1} + \dfrac{|\psi_2|^2}{2m_2} + 2\eta|\psi_1||\psi_2|\cos\phi\right)\dfrac{d^2\theta}{dx^2} - 2\eta|\psi_1||\psi_2|\dfrac{d\phi}{dx}\dfrac{d\theta}{dx}\sin\phi = 0 \\ 2\dfrac{d}{dx}\left(C(\phi)\dfrac{d\phi}{dx}\right) - \dfrac{dB(\phi)}{d\phi}\left(\dfrac{d\theta}{dx}\right)^2 - \dfrac{dC(\phi)}{d\phi}\left(\dfrac{d\phi}{dx}\right)^2 - D\sin\phi = 0 \end{cases} \quad (A8)$$

In particular case $\eta = 0$ (no drag effect) (A8) coincides with obtained in Ref. [33].